\documentclass{PoS}

\title{Secondary Fe-peak nuclei in the Tycho Supernova Remnant: A Promising Tracer of Type Ia Progenitor Metallicity}

\ShortTitle{Secondary Fe-peak Nuclei in Tycho}

\author{\speaker{Carles Badenes}\thanks{\textit{Chandra} Fellow.}\\
        Princeton University\\
        E-mail: \email{badenes@astro.princeton.edu}}

\author{Eduardo Bravo\\
        Universitat Polit\`ecnica de Catalunya and Institut d'Estudis Espacials de Catalunya\\
        E-mail: \email{eduardo.bravo@upc.edu}}

\author{John P. Hughes\\
        Rutgers University\\
        E-mail: \email{jph@physics.rutgers.edu}}

      \abstract{The Mn to Cr mass ratio in supernova ejecta has recently been proposed as a tracer of Type Ia SN
        progenitor metallicity. We review the advantages and problems of this observable quantity, and discuss them in
        the framework of two Galactic supernova remnants: the well known Tycho SNR and W49B, an older object that has
        been tentatively classified as Type Ia. The fluxes of the Mn and Cr K$\alpha$ lines in the X-ray spectra of
        these SNRs observed by the \textit{Suzaku} and \textit{ASCA} satellites suggest progenitors of supersolar
        metallicity for both objects.}

\FullConference{10th Symposium on Nuclei in the Cosmos\\
		 July 27 - August 1 2008\\
		 Mackinac Island, Michigan, USA}

\begin{document}

\section{Mn/Cr as a Tracer of Metallicity in Type Ia SN Progenitors}

The recent detection of the Cr and Mn K$\alpha$ lines in the X-ray spectrum of the Tycho Supernova Remnant (SNR) by the
Japanese satellite \textit{Suzaku} \cite{tamagawa08:tycho} has opened the possibility to study secondary Fe-peak nuclei
in the shocked ejecta of Type Ia SNRs. The presence of Mn in particular, which has an odd
atomic number, has important implications for constraining several properties of Type Ia progenitors, as described by
Badenes et al. in \cite{badenes08:mntocr}. Under the appropriate conditions, the Mn to Cr mass ratio in the ejecta of
Type Ia SNe ($M_{Mn}/M_{Cr}$) can be used to calculate the metallicity of the SN progenitor. In these Proceedings, we
expand on the information presented in \cite{badenes08:mntocr}, providing more details on the models, the advantages and
limitations of $M_{Mn}/M_{Cr}$ as a tracer of progenitor metallicity, and the applicability of the method to
current and future X-ray observations of SNRs.

The rationale for the use of $M_{Mn}/M_{Cr}$ as a tracer of progenitor metallicity $Z$ was explained in detail in
\cite{badenes08:mntocr} \footnote{Here and in \cite{badenes08:mntocr}, we define $Z$ as the mass fraction of all
  elements heavier than He. Note that the correspondence between this theoretical quantity and a given observational
  tracer of metallicity (like $\mathrm{[Fe/H]}$, for instance) is not necessarily trivial.}. It is based on an argument
by Timmes et al. \cite{timmes03:variations_peak_luminosity_SNIa} that connects the amount of C, N, and O in the
progenitor star (and hence its $Z$) to the trace amount of $^{22}$Ne that is present in the CO white dwarf (WD) that
will eventually explode as a SN Ia. During the explosion itself, both $^{55}$Co and $^{52}$Fe (the parent nuclei of
$^{55}$Mn and $^{52}$Cr, respectively) are synthesized in the incomplete Si burning regime. These nuclides belong to a
quasi-statistical equilibrium group dominated by $^{56}$Ni. While $^{52}$Fe is linked to $^{56}$Ni by the reaction
$\mathrm{^{52}Fe(\alpha,\gamma)^{56}Ni}$, which is not sensitive to $\eta$, $^{55}$Co is linked to $^{56}$Ni by the
$\mathrm{^{55}Co(p,\gamma)^{56}Ni}$ reaction. In this last reaction, the proton abundance is strongly dependent on
$\eta$, with larger values of $\eta$ leading to lower proton abundances, which favors the synthesis of $^{55}$Co in
progenitors with high metallicity. In this context, it is worth mentioning the work of \cite{cescutti08:Mn_evolution},
who find evidence for a metallicity dependent yield of Mn in SN Ia, with Mn synthesis appearing enhanced in high
metallicity stellar systems like the Galactic bulge.

In \cite{badenes08:mntocr}, the nucleosynthetic output of 36 Type Ia SN models calculated with different trace amounts
of $^{22}$Ne in the CO WD was examined to quantify the relationship between $Z$ and $M_{Mn}/M_{Cr}$ (Figure
\ref{fig1}). In these calculations, the inner 0.2 $\mathrm{M_{\odot}}$ of ejecta were not included in the final Mn/Cr
ratio. Inside this region, neutron-rich nuclear statistical equilibrium (NSE) takes place, and minor quantities of Mn
and Cr are produced at a mass ratio that is independent of the value of $Z$. Removal of the n-rich NSE products is
justified by the fact that the reverse shock in most ejecta-dominated Type Ia SNRs in our Galaxy has not reached the
inner 0.2 $\mathrm{M_{\odot}}$ of ejecta, and this material does not appear to mix into the outer layers either during
the SN phase \cite{mazzali07:zorro} or the SNR phase \cite{fesen07:SN1885}. Another argument that supports the exclusion
of the n-rich NSE products from an observational point of view is the absence of the K$\alpha$ line from Ni in the same
exposure of the Tycho SNR that revealed the Mn and Cr lines \cite{tamagawa08:tycho}. If a large amount of n-rich NSE
material (or for that matter, of any kind of NSE material) had been thermalized by the reverse shock, this line would
show up at 7.5 keV in the \textit{Suzaku} spectrum. Fitting a power law to the points shown in Figure \ref{fig1} yields
the relation $M_{Mn}/M_{Cr}=5.3 \times Z^{0.65}$ \cite{badenes08:mntocr}, which is virtually independent of the details
of the explosion dynamics.

\section{The Impact of C simmering on the Mn/Cr ratio}

Before a slowly accreting WD explodes as a SN Ia, there is a $\sim$ 1000 yr long phase of slow C fusion in its core. The
energy input from this `simmering' creates a more or less extended convective region inside the WD. It was pointed out
by \cite{piro08:neutronization_SNIasimmering} that the weak interactions which take place during this simmering phase can
increase the value of $\eta$ in the WD material, albeit only by $\Delta \eta=0.0015$
\cite{chamulak08:reduction_electron_simmering_SNIa}. The impact that this increase of $\eta$ will have on the
$M_{Mn}/M_{Cr}$ ratio will depend on the extent of the overlap between the convective core and the explosive Si burning
region of the ejecta where Mn and Cr are synthesized. Unfortunately, the extent of the convective core in a
pre-explosion CO WD is not known, and cannot be calculated self-consistently without sophisticated simulations
\cite{piro08:neutronization_SNIasimmering2}.

To estimate this impact, we consider two limiting cases. If the entire WD is convective, all the explosive Si burning
products are affected by simmering, and the $M_{Mn}/M_{Cr}$ ratio has a lower bound of 0.4, which does not allow to
measure metallicities below solar (see Figure \ref{fig1}). Another possibility is that the size of the convective core
is limited by the Ledoux criterion to the central C-depleted region created during hydrostatic He-burning, as proposed
by \cite{hoeflich02:runaway}. There are a number of reasons why this seems plausible (see \cite{badenes08:mntocr}), but
there is no way to prove that it is indeed the case. In this scenario, the impact of simmering is only large for
subluminous Type Ia SNe, whose Si-rich regions reach deeper into the SN ejecta. The effect is also stronger at lower
$Z$, because metal-poor stars have larger C-depleted cores \cite{dominguez01:SNIa_progenitors}. Two examples of
simmering-modified models with limited convection are shown in Figure \ref{fig1}. Under these conditions, the impact of
C simmering should be of no concern for the Tycho SNR, because we know both from the historical light curve
\cite{ruiz-lapuente04:TychoSN} and the X-ray emission of the SNR \cite{badenes06:tycho} that the SN of 1572 was not
subluminous, but rather normal or slightly overluminous.

\section{Observations: Tycho, W49B and Beyond}

\begin{figure}
  \begin{center}
    \includegraphics[width=.3\textwidth,angle=90]{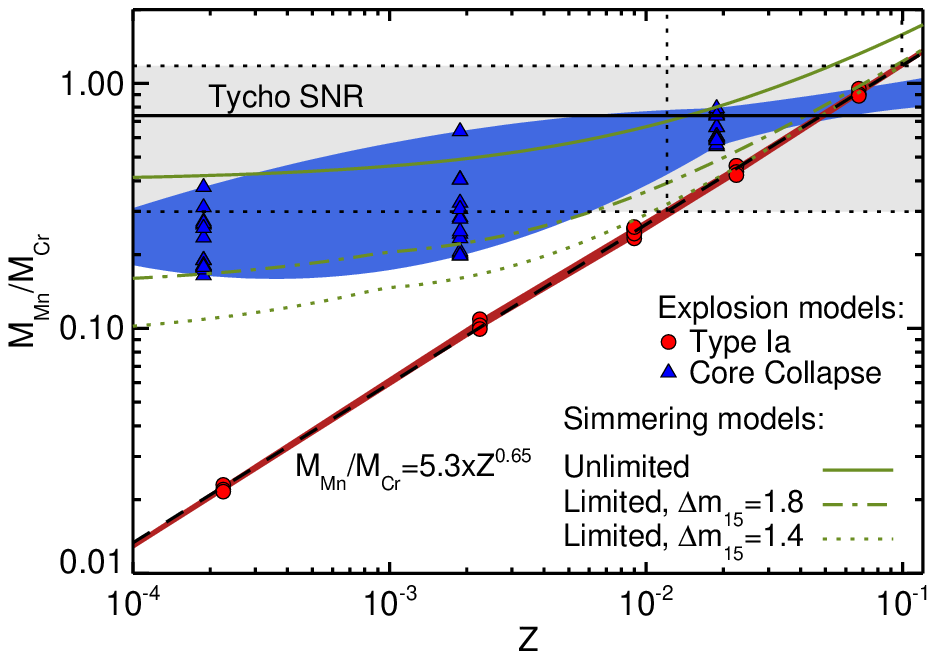}
    \includegraphics[width=.3\textwidth,angle=90]{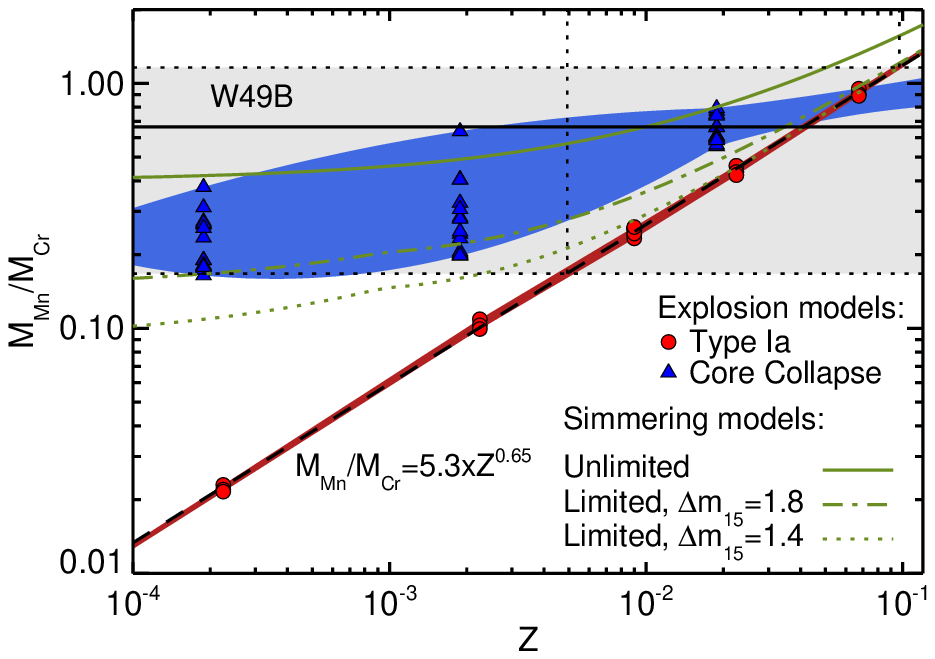}
    \caption{The $M_{Mn}/M_{Cr}$ ratio vs. $Z$ in Type Ia (red circles) and core collapse (blue triangles) SN models
      \cite{badenes08:mntocr,woosley95:core_collapse_models}, together with the observed $M_{Mn}/M_{Cr}$ for Tycho (left
      panel, adapted from \cite{badenes08:mntocr}) and W49B (right panel, calculated from the line fluxes given in
      \cite{hwang00:W49B}).  Some simple models for the impact of simmering are also shown. The solid green plot
      represents the most pessimistic case of a fully convective WD. The dash-dotted and dotted green plots represent
      the more optimistic case of a convective core limited to the C-depleted region of the WD, for a subluminous
      ($\Delta m_{15}=1.8$), and a normal, but faint ($\Delta m_{15}=1.4$) Type Ia SN. In the models with limited
      convection, it has been assumed that the WD progenitor had a ZAMS mass of $5\,\mathrm{M_{\odot}}$.}
    \label{fig1}
  \end{center}
\end{figure}

Assuming no impact from C simmering, the metallicity of the Tycho SN progenitor would be $Z=0.048^{+0.051}_{-0.036}$
(from \cite{badenes08:mntocr}, see Figure \ref{fig1}). Here we also present the case of the Galactic SNR W49B, another
object which has Mn and Cr lines in its X-ray spectrum \cite{hwang00:W49B}, but whose Type Ia origin and age are
uncertain (see discussion in \cite{badenes07:outflows}). In this case, the measured $M_{Mn}/M_{Cr}=0.66\pm0.50$
translates into $Z=0.041^{+0.056}_{-0.036}$, also without simmering impact (Figure \ref{fig1}). This metallicity value
should be considered with caution, because (a) the SN that originated SNR W49B might have been subluminous, making the
impact of simmering important even in the case of limited convection; and (b) given the unknown age, the reverse shock
may have propagated into the n-rich NSE region, which would contaminate the measured $M_{Mn}/M_{Cr}$.

These two test cases can serve to illustrate several points. First, both Tycho and W49B have a larger $M_{Mn}/M_{Cr}$
that can be explained by C simmering alone, which suggests a solar or supersolar metallicity for both events, regardless
of the size of the convective core. Second, the large uncertainties in the measured values of $M_{Mn}/M_{Cr}$ make it
very difficult to completely discard \textit{some} impact from simmering, specially in the most pessimistic case of
unlimited convection. This underlines the importance of improving the $M_{Mn}/M_{Cr}$ measurements, both by using
adequate atomic data for Mn and Cr (in \cite{badenes08:mntocr}, the specific emissivities had to be interpolated, which
introduced a large uncertainty in the measurement) and by calculating line fluxes from deeper observations that can
reduce the statistical errors. Determining accurate fluxes for such weak lines represents a challenge for the present
generation of X-ray telescopes. An improvement on the measurements presented here is certainly possible for bright SNRs,
but more distant objects will require either much larger collecting areas or a new type of detectors, such as the
microcalorimeters being planned for the \textit{NeXT/ASTRO-H} mission \cite{fujimoto02:calorimeter}.

\section{Conclusions}

In \cite{badenes08:mntocr}, the $M_{Mn}/M_{Cr}$ ratio was introduced as a tracer of SN Ia progenitor metallicity. It
might be more appropriate to say that it is a tracer of \textit{neutron excess} in the explosive Si burning region of
Type Ia SNe. This is interesting in its own right, because it also opens a window into the extent of the convective core
of pre-explosion WDs. We hope that more and better observations will let us disentangle the contributions of metallicity
and C simmering to the neutron excess in Type Ia SNe, and that this can help us to understand the lingering mystery of
Type Ia progenitors.

\acknowledgments{Support for this work was provided by NASA through Chandra Postdoctoral Fellowship Award Number
  PF6-70046 issued by the Chandra X-ray Observatory Center, which is operated by the Smithsonian Astrophysical
  Observatory for and on behalf of NASA under contract NAS8-03060. EB is supported by grants AYA2007-66256 and
  AYA2005-08013-C03-01. JPH is partially supported by NASA grant NNG05GP87G.}

\end{document}